**Canonical Momenta Indicators of Financial Markets and Neocortical EEG**

Lester Ingber
Lester Ingber Research
P.O. Box 857, McLean, Virginia 22101, U.S.A.
ingber@ingber.com, ingber@alumni.caltech.edu

**Abstract**—A paradigm of statistical mechanics of financial markets (SMFM) is fit to multivariate financial markets using Adaptive Simulated Annealing (ASA), a global optimization algorithm, to perform maximum likelihood fits of Lagrangians defined by path integrals of multivariate conditional probabilities. Canonical momenta are thereby derived and used as technical indicators in a recursive ASA optimization process to tune trading rules. These trading rules are then used on out-of-sample data, to demonstrate that they can profit from the SMFM model, to illustrate that these markets are likely not efficient. This methodology can be extended to other systems, e.g., electroencephalography. This approach to complex systems emphasizes the utility of blending an intuitive and powerful mathematical-physics formalism to generate indicators which are used by AI-type rule-based models of management.

## 1. Introduction

Over a decade ago, the author published a paper suggesting the use of newly developed methods of multivariate nonlinear nonequilibrium calculus to approach a statistical mechanics of financial markets (SMFM) [1]. These methods were applied to interest-rate term-structure systems [2,3]. Still, for some time, the standard accepted paradigm of financial markets has been rooted in equilibrium processes [4]. There is a current effort by many to examine nonlinear and nonequilibrium processes in these markets [5], and this paper reinforces this point of view. Another paper gives some earlier 1991 results using this approach [6].

There are several issues that are clarified here, by presenting calculations of a specific trading model: (A) It is demonstrated how multivariate markets might be formulated in a nonequilibrium paradigm. (B) It is demonstrated that numerical methods of global optimization can be used to fit such SMFM models to data. (C) A variational principle possessed by SMFM permits derivation of technical indicators, such as canonical momenta, that can be used to describe deviations from most likely evolving states of the multivariate system. (D) These technical indicators can be embedded in realistic trading scenarios, to test whether they can profit from nonequilibrium in markets.

Section 2 outlines the formalism used to develop the nonlinear nonequilibrium SMFM model. Section 3 describes application of SMFM to SP500 cash and future data, using Adaptive Simulated Annealing (ASA) [7] to fit the short-time conditional probabilities developed in Section 2, and to establish trading rules by recursively optimizing with ASA, using optimized technical indicators developed from SMFM. These calculations were briefly mentioned in another ASA paper [8]. Section 4 describes similar applications, now in progress, to correlating customized electroencephalographic (EEG) momenta indicators to physiological and behavioral states of humans. Section 5 is a brief conclusion.

## 2. SMFM Model

### 2.1. Random walk model

The use of Brownian motion as a model for financial systems is generally attributed to Bachelier [9], though he incorrectly intuited that the noise scaled linearly instead of as the square root relative to the random log-price variable. Einstein is generally credited with using the correct mathematical description in a larger physical context of statistical systems. However, several studies imply that changing prices of many markets do not follow a random walk, that they may have long-term dependences in price correlations, and that they may not be efficient in quickly arbitraging new information [10-12]. A random walk for returns, rate of change of prices over prices, is described by a Langevin equation with simple additive noise $\eta$, typically representing the continual random influx of information into the market.

$$\dot{\Gamma} = -\gamma_1 + \gamma_2 \eta ,$$

$$\dot{\Gamma} = d\Gamma/dt ,$$

$$< \eta(t) >_\eta = 0 , \quad < \eta(t), \eta(t') >_\eta = \delta(t - t') , \tag{1}$$

where $\gamma_1$ and $\gamma_2$ are constants, and $\Gamma$ is the logarithm of (scaled) price. Price, although the most dramatic observable, may not be the only appropriate dependent variable or order parameter for the system of markets [13]. This possibility has also been called the "semistrong form of the efficient market hypothesis" [10].

It is necessary to explore the possibilities that a given market evolves in nonequilibrium, e.g., evolving irreversibly, as well as nonlinearly, e.g., $\gamma_{1,2}$ may be functions of $\Gamma$. Irreversibility, e.g., causality [14] and nonlinearity [15], have been suggested as processes necessary to take into account in order to understand markets, but modern methods of statistical mechanics now provide a more explicit paradigm to consistently include these processes in *bona fide* probability distributions. Reservations have been expressed about these earlier models at the time of their presentation [16].

Developments in nonlinear nonequilibrium statistical mechanics in the late 1970's and their application to a variety of testable physical phenomena illustrate the importance of properly treating nonlinearities and nonequilibrium in systems where simpler analyses prototypical of linear equilibrium Brownian motion do not suffice [17].

### 2.2. Statistical mechanics of large systems

Aggregation problems in nonlinear nonequilibrium systems, e.g., as defines a market composed of many traders [1], typically are "solved" (accommodated) by having new entities/languages developed at these disparate scales in order to efficiently pass information back and forth [18,19]. This is quite different from the nature of quasi-equilibrium quasi-linear systems, where thermodynamic or cybernetic approaches are possible. These approaches typically fail for nonequilibrium nonlinear systems.

These new methods of nonlinear statistical mechanics only recently have been applied to complex large-scale physical problems, demonstrating that observed data can be described by the use of these algebraic functional forms. Success was gained for large-scale systems in neuroscience, in a series of papers on statistical mechanics of neocortical interactions [20-30], and in nuclear physics [31-33]. This methodology has been used for problems in combat analyses [19,34-37]. These methods have been suggested for financial markets [1], applied to a term structure model of interest rates [2,3], and to optimization of trading [6].

### 2.3. Statistical development

When other order parameters in addition to price are included to study markets, Eq. (1) is accordingly generalized to a set of Langevin equations.

$$\dot{M}^G = f^G + \hat{g}_j^G \eta^j \, , \, (G = 1, \cdots, \Lambda) \, , \, (j = 1, \cdots, N) \, ,$$

$$\dot{M}^G = dM^G/d\Theta \, ,$$

$$< \eta^j(\Theta) >_\eta = 0 \, , \, < \eta^j(\Theta), \eta^{j'}(\Theta') >_\eta = \delta^{jj'} \delta(\Theta - \Theta') \, , \tag{2}$$

where $f^G$ and $\hat{g}_j^G$ are generally nonlinear functions of mesoscopic order parameters $M^G$, $j$ is a microscopic index indicating the source of fluctuations, and $N \geq \Lambda$. The Einstein convention of summing over repeated indices is used. Vertical bars on an index, e.g., $|j|$, imply no sum is to be taken on repeated indices. $\Theta$ is used here to emphasize that the most appropriate time scale for trading may not be real time $t$.

Via a somewhat lengthy, albeit instructive calculation, outlined in several other papers [1,3,25], involving an intermediate derivation of a corresponding Fokker-Planck or Schrödinger-type equation for the conditional probability distribution $P[M(\Theta)|M(\Theta_0)]$, the Langevin rate Eq. (2) is developed into the probability distribution for $M^G$ at long-time macroscopic time event $\Theta = (u+1)\theta + \Theta_0$, in terms of a Stratonovich path-integral over mesoscopic Gaussian conditional probabilities [38-40]. Here, macroscopic variables are defined as the long-time limit of the evolving mesoscopic system. The corresponding Schrödinger-type equation is [39,41]

$$\partial P/\partial \Theta = \frac{1}{2}(g^{GG'}P)_{,GG'} - (g^G P)_{,G} + V \, ,$$

$$g^{GG'} = k_T \delta^{jk} \hat{g}_j^G \hat{g}_k^{G'} \, , \, g^G = f^G + \frac{1}{2} \delta^{jk} \hat{g}_j^{G'} \hat{g}_{k,G'}^G \, ,$$

$$[\cdots]_{,G} = \partial[\cdots]/\partial M^G \, . \tag{3}$$

This is properly referred to as a Fokker-Planck equation when $V \equiv 0$. Note that although the partial differential Eq. (3) contains equivalent information regarding $M^G$ as in the stochastic differential Eq. (2), all references to $j$ have been properly averaged over. I.e., $\hat{g}_j^G$ in Eq. (2) is an entity with parameters in both microscopic and mesoscopic spaces, but $\hat{M}$ is a purely mesoscopic variable, and this is more clearly reflected in Eq. (3).

The path integral representation is given in terms of the Lagrangian $L$.

$$P[M_\Theta|M_{\Theta_0}]dM(\Theta) = \int \cdots \int DM \exp(-S)\delta[M(\Theta_0) = M_0]\delta[M(\Theta) = M_\Theta] \, ,$$

$$S = k_T^{-1} \min \int_{\Theta_0}^{\Theta} d\Theta' L \, ,$$

$$DM = \lim_{u\to\infty} \prod_{\rho=1}^{u+1} g^{1/2} \prod_G (2\pi\theta)^{-1/2} dM_\rho^G ,$$

$$L(\dot{M}^G, M^G, \Theta) = \frac{1}{2}(\dot{M}^G - h^G)g_{GG'}(\dot{M}^{G'} - h^{G'}) + \frac{1}{2} h^G_{;G} + R/6 - V ,$$

$$h^G = g^G - \frac{1}{2} g^{-1/2}(g^{1/2} g^{GG'})_{,G'} ,$$

$$g_{GG'} = (g^{GG'})^{-1} , \quad g = \det(g_{GG'}) ,$$

$$h^G_{;G} = h^G_{,G} + \Gamma^F_{GF} h^G = g^{-1/2}(g^{1/2} h^G)_{,G} ,$$

$$\Gamma^F_{JK} \equiv g^{LF}[JK, L] = g^{LF}(g_{JL,K} + g_{KL,J} - g_{JK,L}) ,$$

$$R = g^{JL} R_{JL} = g^{JL} g^{JK} R_{FJKL} ,$$

$$R_{FJKL} = \frac{1}{2}(g_{FK,JL} - g_{JK,FL} - g_{FL,JK} + g_{JL,FK}) + g_{MN}(\Gamma^M_{FK}\Gamma^N_{JL} - \Gamma^M_{FL}\Gamma^N_{JK}) . \tag{4}$$

Mesoscopic variables have been defined as $M^G$ in the Langevin and Fokker-Planck representations, in terms of their development from the microscopic system labeled by $j$. The Riemannian curvature term $R$ arises from nonlinear $g_{GG'}$, which is a bona fide metric of this parameter space [39].

### 2.4. Algebraic complexity yields simple intuitive results

It must be emphasized that the output need not be confined to complex algebraic forms or tables of numbers. Because $L$ possesses a variational principle, sets of contour graphs, at different long-time epochs of the path-integral of $P$ over its variables at all intermediate times, give a visually intuitive and accurate decision-aid to view the dynamic evolution of the scenario. For example, this Lagrangian approach permits a quantitative assessment of concepts usually only loosely defined.

$$\text{``Momentum''} = \Pi^G = \frac{\partial L}{\partial(\partial M^G/\partial\Theta)} ,$$

$$\text{``Mass''} g_{GG'} = \frac{\partial^2 L}{\partial(\partial M^G/\partial\Theta)\partial(\partial M^{G'}/\partial\Theta)} ,$$

$$\text{``Force''} = \frac{\partial L}{\partial M^G} ,$$

$$\text{``}F = ma\text{''}: \quad \delta L = 0 = \frac{\partial L}{\partial M^G} - \frac{\partial}{\partial\Theta}\frac{\partial L}{\partial(\partial M^G/\partial\Theta)} , \tag{5}$$

where $M^G$ are the variables and $L$ is the Lagrangian. These physical entities provide another form of intuitive, but quantitatively precise, presentation of these analyses. For example, daily newspapers use this terminology to discuss the movement of security prices. Here, we will use the canonical momenta as indicators to develop trading rules.

### 2.5. Fitting parameters

The short-time path-integral Lagrangian of a $\Lambda$-dimensional system can be developed into a scalar "dynamic cost function," $C$, in terms of parameters, e.g., generically represented as $C(\tilde{\alpha})$,

$$C(\tilde{\alpha}) = L\Delta\Theta + \frac{\Lambda}{2} \ln(2\pi\Delta\Theta) - \frac{1}{2} \ln g , \tag{6}$$

which can be used with the ASA algorithm [7], originally called Very Fast Simulated Reannealing (VFSR) [42], to find the (statistically) best fit of parameters. The cost function for a given system is obtained by the product of $P$'s over all data epochs, i.e., a sum of $C$'s is obtained. Then, since we essentially are performing a maximum likelihood fit, the cost functions obtained from somewhat different theories or data can provide a relative statistical measure of their likelihood, e.g., $P_{12} \sim \exp(C_2 - C_1)$.

If there are competing mathematical forms, then it is advantageous to utilize the path-integral to calculate the long-time evolution of $P$ [19,35]. Experience has demonstrated that the long-time correlations derived from theory, measured against the observed data, is a viable and expedient way of rejecting models not in accord with observed evidence.

## 2.6. Numerical methodology

ASA [42] fits short-time probability distributions to observed data, using a maximum likelihood technique on the Lagrangian. This algorithm has been developed to fit observed data to a theoretical cost function over a $D$-dimensional parameter space [42], adapting for varying sensitivities of parameters during the fit.

Simulated annealing (SA) was developed in 1983 to deal with highly nonlinear problems [43], as an extension of a Monte-Carlo importance-sampling technique developed in 1953 for chemical physics problems. It helps to visualize the problems presented by such complex systems as a geographical terrain. For example, consider a mountain range, with two "parameters," e.g., along the North–South and East–West directions. We wish to find the lowest valley in this terrain. SA approaches this problem similar to using a bouncing ball that can bounce over mountains from valley to valley. We start at a high "temperature," where the temperature is an SA parameter that mimics the effect of a fast moving particle in a hot object like a hot molten metal, thereby permitting the ball to make very high bounces and being able to bounce over any mountain to access any valley, given enough bounces. As the temperature is made relatively colder, the ball cannot bounce so high, and it also can settle to become trapped in relatively smaller ranges of valleys.

We imagine that our mountain range is aptly described by a "cost function." We define probability distributions of the two directional parameters, called generating distributions since they generate possible valleys or states we are to explore. We define another distribution, called the acceptance distribution, which depends on the difference of cost functions of the present generated valley we are to explore and the last saved lowest valley. The acceptance distribution decides probabilistically whether to stay in a new lower valley or to bounce out of it. All the generating and acceptance distributions depend on temperatures.

In 1984 [44], it was established that SA possessed a proof that, by carefully controlling the rates of cooling of temperatures, it could statistically find the best minimum, e.g., the lowest valley of our example above. This was good news for people trying to solve hard problems which could not be solved by other algorithms. The bad news was that the guarantee was only good if they were willing to run SA forever. In 1987, a method of fast annealing (FA) was developed [45], which permitted lowering the temperature exponentially faster, thereby statistically guaranteeing that the minimum could be found in some finite time. However, that time still could be quite long. Shortly thereafter, in 1987 the author developed Very Fast Simulated Reannealing (VFSR) [42], now called Adaptive Simulated Annealing (ASA), which is exponentially faster than FA. It is used world-wide across many disciplines [8], and the feedback of many users regularly scrutinizing the source code ensures the soundness of the code as it becomes more flexible and powerful [46].

ASA has been applied to many problems by many people in many disciplines [8,46,47]. The code is available via anonymous ftp from ftp.ingber.com, which also can be accessed via the world-wide web (WWW) as http://www.ingber.com/.

## 3. Fitting SMFM to SP500

### 3.1. Data processing

For the purposes of this paper, it suffices to consider a two-variable problem, SP500 prices of futures, $p^1$, and cash, $p^2$. (Note that in a previous paper [6], these two variables were inadvertently incorrectly reversed.) Data included 251 points of 1989 and 252 points of 1990 daily closing data. Time between data was taken as real time $t$, e.g., a weekend added two days to the time between data of a Monday and a previous Friday.

It was decided that relative data should be more important to the dynamics of the SMFM model than absolute data, and an arbitrary form was developed to preprocess data used in the fits,

$$M^i(t) = p^i(t + \Delta t)/p^i(t) \, , \tag{7}$$

where $i = \{1, 2\}$ = {futures, cash}, and $\Delta t$ was the time between neighboring data points, and $t + \Delta t$ is the current trading time. The ratio served to served to suppress strong drifts in the absolute data.

### 3.2. ASA fits of SMFM to data

Two source of noise were assumed, so that the equations of this SMFM model are

$$\frac{dM^G}{dt} = \sum_{G'=1}^{2} f_{G'}^G M^{G'} + \sum_{i=1}^{2} \hat{g}_i^G \eta^i \, , \ G = \{1, 2\} \, . \tag{8}$$

The 8 parameters, $\{f_{G'}^G, \hat{g}_i^G\}$ were all taken to be constants.

As discussed previously, the path-integral representation was used to define an effective cost function. Minimization of the cost function was performed using ASA. Some experimentation with the fitting process led to a scheme whereby after sufficient importance-sampling, the optimization was shunted over to a quasi-local code, the Broyden-Fletcher-Goldfarb-Shanno (BFGS) algorithm [48], to add another decimal of precision. If ASA was shunted over too quickly to BFGS, then poor fits were obtained, i.e., the fit stopped in a higher local minimum.

Using 1989 data, the parameters $f_{G'}^{G}$ were constrained to lie between -1.0 and 1.0. The parameters $\hat{g}_i^G$ were constrained to lie between 0 and 1.0. The values of the parameters, obtained by this fitting process were: $f_1^1 = 0.0686821$, $f_2^1 = -0.068713$, $\hat{g}_1^1 = 0.000122309$, $\hat{g}_2^1 = 0.000224755$, $f_1^2 = 0.645019$, $f_2^2 = -0.645172$, $\hat{g}_1^2 = 0.00209127$, $\hat{g}_2^2 = 0.00122221$.

### 3.3. ASA fits of trading rules

A simple model of trading was developed. Two time-weighted moving averages, of wide and narrow windows, $a_w$ and $a_n$ were defined for each of the two momenta variables. During each new epoch of $a_w$, always using the fits of the SMFM model described in the previous section as a zeroth order estimate, the parameters $\{f_{G'}^{G}, \hat{g}_i^G\}$ were refit using data within each epoch. Averaged canonical momenta, i.e., using Eq. (5), were calculated for each new set of $a_w$ and $a_n$ windows. Fluctuation parameters $\Delta\Pi_w^G$ and $\Delta\Pi_n^G$, were defined, such that any change in trading position required that there was some reasonable information outside of these fluctuations that could be used as criteria for trading decisions. No trading was performed for the first few days of the year until the momenta could be calculated. Commissions of $70 were paid every time a new trade of 100 units was taken. Thus, there were 6 trading parameters used in this example, $\{a_w, a_n, \Delta\Pi_w^G, \Delta\Pi_n^G\}$.

The order of choices made for daily trading are as follows. A 0 represents no positions are open and no trading is performed until enough data is gathered, e.g., to calculate momenta. A 1 represents entering a long position, whether from a waiting or a short position, or a current long position was maintained. This was performed if the both wide-window and narrow-window averaged momenta of both cash and futures prices were both greater than their $\Delta\Pi_w^G$ and $\Delta\Pi_n^G$ fluctuation parameters. A −1 represents entering a short position, whether from a waiting or a long position, or a current short position was maintained. This was performed if the both wide-window and narrow-window averaged momenta of both cash and futures prices were both less than their $\Delta\Pi_w^G$ and $\Delta\Pi_n^G$ fluctuation parameters.

### 3.4. In-sample ASA fits of trading rules

For the data of 1989, recursive optimization was performed. The trading parameters were optimized in an outer shell, using the negative of the net yearly profit/loss as a cost function. This could have been weighted by something like the absolute value of maximum loss to help minimize risk, but this was not done here. The inner shell of optimization fine-tuning of the SMFM model was performed daily over the current $a_w$ epoch.

At first, ASA and shunting over to BFGS was used for each shell, but it was realized that good results could be obtained using ASA and BFGS on the outer shell, and just BFGS on the inner shell (always using the ASA and BFGS derived zeroth order SMFM parameters as described above). Thus, recursive optimization was performed to establish the required goodness-of-fit, and more efficient local optimization was used only in those instances where it could replicate the global optimization. This is expected to be quite system dependent.

The trading-rule parameters were constrained to lie within the following ranges: $a_w$ integers between 15 and 25, $a_n$ integers between 3 and 14, $\Delta\Pi_w^G$ and $\Delta\Pi_n^G$ between 0 and 200. The trading parameters fit by this procedure were: $a_w = 18$, $a_n = 11$, $\Delta\Pi_w^1 = 30.3474$, $\Delta\Pi_w^2 = 98.0307$, $\Delta\Pi_n^1 = 11.2855$, $\Delta\Pi_n^2 = 54.8492$.

The summary of results was: cumulative profit = $54170, number of profitable long positions = 11, number of profitable short positions = 8, number of losing long positions = 5, number of losing short positions = 6, maximum profit of any given trade = $11005, maximum loss of any trade = –$2545, maximum accumulated profit during year = $54170, maximum loss sustained during year = $0.

### 3.5. Out-of-sample SMFM trading

The trading process described above was applied to the 1990 out-of-sample SP500 data. Note that 1990 was a "bear" market, while 1989 was a "bull" market. Thus, these two years had quite different overall contexts, and this was believed to provide a stronger test of this methodology than picking two years with similar contexts.

The inner shell of optimization was performed as described above for 1990 as well. The summary of results was: cumulative profit = $28300, number of profitable long positions = 10, number of profitable short positions = 6, number of losing long positions = 6, number of losing short positions = 10, maximum profit of any given trade = $6780, maximum loss of any trade = –$2450, maximum accumulated profit during year = $29965, maximum loss sustained during year = –$5945. Tables of results are available as file markets96_momenta_tbl.txt.Z in http://www.ingber.com/MISC.DIR/ and ftp.ingber.com/MISC.DIR.

Only one variable, the futures SP500, was actually traded, albeit the code can accommodate trading on multiple markets. There is more leverage and liquidity in actually trading the futures market. The multivariable coupling to the cash market entered in three important ways: (1) The SMFM fits were to the coupled system, requiring a global optimization of all parameters in both markets to define the time evolution of the futures market. (2) The canonical momenta for the futures market is in terms of the partial derivative of the full Lagrangian; the dependency on the cash market enters both as a function of the relative value of the off-diagonal to diagonal terms in the metric, as well as a contribution to the drifts and diffusions from this market. (3) The canonical momenta of both markets were used as technical indicators for trading the futures market.

### 3.6. Reversing data sets

The same procedures described above were repeated, but using the 1990 SP500 data set for training and the 1989 data set for testing.

For the training phase, using 1990 data, the parameters $f_{G'}^G$ were constrained to lie between -1.0 and 1.0. The parameters $\hat{g}_i^G$ were constrained to lie between 0 and 1.0. The values of the parameters, obtained by this fitting process were: $f_1^1 = 0.0685466$, $f_2^1 = -0.068571$, $\hat{g}_1^1 = 7.52368\ 10^{-6}$, $\hat{g}_2^1 = 0.000274467$, $f_1^2 = 0.642585$, $f_2^2 = -0.642732$, $\hat{g}_1^2 = 9.30768\ 10^{-5}$, $\hat{g}_2^2 = 0.00265532$. Note that these values are quite close to those obtained above when fitting the 1989 data.

The trading-rule parameters were constrained to lie within the following ranges: $a_w$ integers between 15 and 25, $a_n$ integers between 3 and 14, $\Delta\Pi_w^G$ and $\Delta\Pi_n^G$ between 0 and 200. The trading parameters fit by this procedure were: $a_w = 11$, $a_n = 8$, $\Delta\Pi_w^1 = 23.2324$, $\Delta\Pi_w^2 = 135.212$, $\Delta\Pi_n^1 = 169.512$, $\Delta\Pi_n^2 = 9.50857$,

The summary of results was: cumulative profit = \$42405, number of profitable long positions = 11, number of profitable short positions = 8, number of losing long positions = 7, number of losing short positions = 6, maximum profit of any given trade = \$8280, maximum loss of any trade = –\$1895, maximum accumulated profit during year = \$47605, maximum loss sustained during year = –\$2915.

For the testing phase, the summary of results was: cumulative profit = \$35790, number of profitable long positions = 10, number of profitable short positions = 6, number of losing long positions = 6, number of losing short positions = 3, maximum profit of any given trade = \$9780, maximum loss of any trade = –\$4270, maximum accumulated profit during year = \$35790, maximum loss sustained during year = \$0. Tables of results are available as file markets96_momenta_tbl.txt.Z in http://www.ingber.com/MISC.DIR/ and ftp.ingber.com/MISC.DIR.

## 4. Extrapolations to EEG

### 4.1. Customized Momenta Indicators of EEG

These techniques are quite generic, and can be applied to a model of statistical mechanics of neocortical interactions (SMNI) which has utilized similar mathematical and numerical algorithms [20-23,25,26,29,30,49]. In this approach, the SMNI model is fit to EEG data, e.g., as previously performed [25]. This develops a zeroth order guess for SMNI parameters for a given subject's training data. Next, ASA is used recursively to seek parameterized predictor rules, e.g., modeled according to guidelines used by clinicians. The parameterized predictor rules form an outer ASA shell, while regularly fine-tuning the SMNI inner-shell parameters within a moving window (one of the outer-shell parameters). The outer-shell cost function is defined as some measure of successful predictions of upcoming EEG events.

In the testing phase, the outer-shell parameters fit in the training phase are used in out-of-sample data. Again, the process of regularly fine-tuning the inner-shell of SMNI parameters is used in this phase.

If these SMNI techniques can find patterns of such such upcoming activity some time before the trained eye of the clinician, then the costs of time and pain in preparation for surgery can be reduced. This project will determine inter-electrode and intra-electrode activities prior to spike activity to determine likely electrode circuitries highly correlated to the onset of seizures. This can only do better than simple averaging or filtering of such activity, as typically used as input to determine dipole locations of activity prior to the onset of seizures.

If a subset of electrode circuitries are determined to be highly correlated to the onset of seizures, then their associated regions of activity can be used as a first approximate of underlying dipole sources of brain activity affecting seizures. This first approximate may be better than using a spherical head model to deduce such a first guess. Such first approximates can then be used for more realistic dipole source modeling, including the actual shape of the brain surface to determine likely localized areas of diseased tissue.

These momenta indicators should be considered as supplemental to other clinical indicators. This is how they are being used in financial trading systems.

## 5. Conclusion

A complete sample scenario has been presented: (a) developing a multivariate nonlinear nonequilibrium model of financial markets; (b) fitting the model to data using methods of ASA global optimization; (c) deriving technical indicators to express dynamics about most likely states; (d) optimizing trading rules using these technical indicators; (e) trading on out-of-sample data to determine if steps (a)–(d) are at least sufficient to profit by the knowledge gained of these financial markets, i.e., these markets are not efficient.

Just based the models and representative calculations presented here, no comparisons can yet be made of any relative superiority of these techniques over other models of markets and other sets of trading rules. Rather, this exercise should be viewed as an explicit demonstration (1) that financial markets can be modeled as nonlinear nonequilibrium systems, and (2) that financial markets are not efficient and that they can be properly fit and profitably traded on real data.

Canonical momenta may offer an intuitive yet detailed coordinate system of some complex systems, which can be used as reasonable indicators of new and/or strong trends of behavior, upon which reasonable decisions and actions can be based. A description has been given of a project in progress, using this same methodology to customize canonical momenta indicators of EEG to human behavioral and physiological states [50].